\documentclass[%
 reprint,
 amsmath,amssymb,
 aps,
]{revtex4-1}

\usepackage{graphicx}
\usepackage{dcolumn}
\usepackage{bm}
\usepackage{longtable}
\usepackage{enumerate}
\usepackage{subfigure}
\usepackage{multirow}
\usepackage{cancel}
\usepackage{tabularx}
\usepackage{threeparttable}
\usepackage{dcolumn}
\usepackage{amssymb}
\usepackage{afterpage}
\usepackage{nicefrac}

\newcommand{\Lower}[1]{\smash{\lower 1.5ex \hbox{#1}}}

\usepackage{ulem}
\newcommand{\down}{\sout{$\downarrow$}}
\newcommand{\up}{\sout{$\uparrow$}}
\newcommand{\zero}{\sout{\phantom{$\downarrow \negthickspace \uparrow$}}}
\newcommand{\double}{\sout{$\downarrow \negthickspace \uparrow$}}

\begin{document}

\preprint{APS/123-QED}
\title{Analysis of two-orbital correlations in wavefunctions restricted to electron-pair states}
\author{Katharina Boguslawski}
\email{k.boguslawski@fizyka.umk.pl}
\author{Pawe{\l} Tecmer}%
\affiliation{%
 Institute of Physics, Faculty of Physics, Astronomy and Informatics, Nicolaus Copernicus University, Grudzi{a}dzka 5, 87-100 Toru{n}, Poland \\
}%
\vspace{-2cm}
 \author{\"Ors Legeza}
\affiliation{%
 Strongly Correlated Systems ``Lend\"ulet" Research Group, Wigner Research Center for Physics, H-1525 Budapest, Hungary
 \\}


\date{\today}

\begin{abstract}
Wavefunctions constructed from electron-pair states can accurately model strong electron correlation effects and are promising approaches especially for larger many-body systems.
In this article, we analyze the nature and the type of electron correlation effects that can be captured by wavefunctions restricted to electron-pair states.
We focus on the Antisymmetric Product of 1-reference orbital Geminal (AP1roG) method combined with an orbital optimization protocol presented in [Phys. Rev. B, 89, 201106(R), 2014] whose performance is assessed against electronic structures obtained form DMRG reference data.
Our numerical analysis covers model systems for strong correlation: the one-dimensional Hubbard model with periodic boundary condition as well as metallic and molecular hydrogen rings. 
Specifically, the accuracy of AP1roG is benchmarked using the single-orbital entropy, the orbital-pair mutual information as well as the eigenvalue spectrum of the one-orbital and two-orbital reduced density matrices.
Our study indicates that contributions from singly occupied states become important in the strong correlation regime which highlights the limitations of the AP1roG method.   
Furthermore, we examine the effect of orbital rotations within the AP1roG model on correlations between orbital pairs.


\end{abstract}

\pacs{Valid PACS appear here}
\maketitle

\section{Introduction}
The many-electron problem remains one of the main challenges of quantum physics and quantum chemistry.
It originates from the fact that electrons do not move independently, but in a correlated fashion.
A quantum description of these effects requires solving the many-body Schr\"odinger equation, for which exact solutions are known only for some model systems~\cite{solvable-models,Hubbard_model,Lieb_1968}.
In practice, we have to rely on approximate methods~\cite{Gus-prx-2015}.
However, many approximate numerical algorithms scale exponentially with system size if the quantum system contains strongly-correlated electrons.
The most promising numerical approaches to treat strongly-correlated fermions are the Density Matrix Renormalization Group (DMRG) algorithm~~\cite{,white,white2,white-qc,scholl05,ors_springer,marti2010b,chanreview,wouters-review,Ors_ijqc,yanai-review,Legeza2015-hubbard} and the quantum Monte Carlo (QMC) method~\cite{QMC,QMC-book-chapter,Stella_H50}.

Another approach to accurately model strongly-correlated electronic systems uses geminals (two-electron basis functions) as building blocks for the electronic wavefunction~\cite{Hurley_1953,Coleman_1965,Silver_1969,Ortiz_1981,Surjan_1999,Kutzelnigg2012,Surjan_2012,Ellis_2013}.
In contrast to geminal-based methods, conventional approaches, like DMRG, exploit one-electron functions (orbitals) to model many-body quantum systems.
In its second-quantized form, a geminal wavefunction can be written as
\begin{equation}
|{\rm Geminal}\rangle =\psi_1^{\dagger}\psi_2^{\dagger}\ldots \psi_{\rm N/2}^{\dagger}|{0} \rangle, 
\end{equation}
with ${N}$ being the number of electrons, $|0 \rangle$ denoting the vacuum state with respect to the creation of the geminals, and $\psi^{\dagger}_{_i}$ is a correlated two-electron function (a geminal). 
If we restrict geminals to be singlet functions, a pair-creation in its natural form, then, reads 
\begin{equation}\label{eq:geminal}
             \psi_i^{\dagger}=\sum_{p=1}^{M_i}c_{p}^{ i}a_{p\uparrow}^{\dagger}a_{p\downarrow}^{\dagger}, 
\end{equation} 
where $M_i$ is the number of one-particle functions (the natural orbitals) used to create geminal $i$, $c_{p}^{ i}$ is a geminal matrix coefficient for subspace $M_i$, and 
$a_{p\uparrow}^{\dagger}$ and $a_{q\downarrow}^{\dagger}$ are the standard electron creation operators for up- and down-spin electrons ($\uparrow$, $\downarrow$).
The structure of the geminal coefficent matrix $\{c_p^i\}$ depends on the geminal wavefunction ansatz used.
Thus, restricting $\{c_p^i\}$, we can derive different flavours for geminal-model wavefunctions~\cite{Limacher_2013}.
The most popular approaches are based on the antisymmetric product of strongly orthogonal geminals~\cite{Hurley_1953,Parr_1956,Parks_1958,Kutzelnigg_1964,Kutzelnigg_1965,Rassolov_2002,Pernal-APSG}, the antisymmetrized geminal power~\cite{Coleman_1965,Coleman_1997,Neuscamman_2012,Eric-orbital-optimization} (which is a special case of projected Hartree--Fock--Bogoliubov~\cite{PHF}), the antisymmetric product of interacting geminals~\cite{Bratoz1965,Silver_1969,Silver_1970,APIG-1,APIG-2,Rosta2002,Surjan_2012} (APIG), or the antisymmetric product of 1-reference-orbital geminals (AP1roG)~\cite{Limacher_2013,OO-AP1roG}. 
Specifically, the AP1roG model allows us to approximate the doubly occupied configuration interaction (DOCI)~\cite{DOCI} wavefunction, but requires only mean-field computational cost in contrast to the factorial scaling of traditional DOCI implementations.
For AP1roG, the sum of eq.~\eqref{eq:geminal} is restricted to run over one occupied orbital $i$ of some reference determinant and all unoccupied orbitals $a$,
\begin{equation}\label{eq:geminal}
             \psi_i^{\dagger}=a_{i\uparrow}^{\dagger}a_{i\downarrow}^{\dagger} + \sum_{a=P+1}^{K}c_{a}^{ i}a_{a\uparrow}^{\dagger}a_{a\downarrow}^{\dagger}, 
\end{equation} 
where $P$ is the number of electron pairs ($P=N/2$ with $N$ being the total number of electrons) and $K$ is the total number of basis functions.

This ansatz for $\psi_i^{\dagger}$ allow us to rewrite the AP1roG wavefunction as a pair-coupled-cluster doubles wavefunction~\cite{p-CCD,Tamar-pCC}, \textit{i.e.},
\begin{equation}\label{eq:ap1rog}
|{\rm AP1roG}\rangle = \exp \left (  \sum_{i=1}^P \sum_{a=P+1}^K c_i^a a_{a \uparrow}^{\dagger}  a_{a\downarrow}^{\dagger} a_{i\downarrow}  a_{i\uparrow}\right )|\Phi_0 \rangle,
\end{equation}
where $|\Phi_0 \rangle$ is some reference determinant. Indices $i$ and $a$ correspond to occupied and virtual sites (orbitals) with respect to $|\Phi_0 \rangle$, $P$ and $K$ again denote the number of electron pairs and orbitals, respectively.
The geminal coefficients $\{c_i^a\}$ thus correspond to the (pair-)coupled-cluster amplitudes.
This wavefunction ansatz is, by construction, size-extensive and has mean-field scaling if the geminal coefficients are optimized using the projected Schr\"odinger equation approach. Note that $|\Phi_0 \rangle$ is optimized as well and hence differs from the Hartree--Fock determinant.

Recent studies demonstrate that AP1roG can reliably model strongly-correlated systems~\cite{pawel_jpca_2014,PS2-AP1roG,Piotrus_Mol-Phys,AP1roG-JCTC,pawel_PCCP2015,Limacher_2015}, even heavy-element containing molecules with multiple degenerate single-particle states~\cite{pawel_PCCP2015}.
However, most of the analysis presented so far was based on energetic arguments or one-body correlation functions like occupation numbers.
In this work, we will present an in-depth analysis of orbital-pair correlations captured by the AP1roG model for one-dimensional systems where quantum fluctuations have a more pronounced role.
Specifically, we will use concepts of quantum information theory to assess orbital entanglement and orbital-pair correlations~\cite{Rissler2006,Barcza_11,Barcza2013,entanglement_letter,entanglement_bonding_2013,Barcza2013,Kasia_ijqc,Ors_ijqc,barcza2014entanglement}, which are particularly instructive to dissect electron correlation effects~\cite{Barcza_11,entanglement_letter}, elucidate chemical reactions~\cite{entanglement_bonding_2013,PCCP_bonding,bonding_qit,boguslawski2014chemical,Roland-RuNO,pawel_PCCP2015,Zhao2015}, and detect changes in the electronic wavefunction~\cite{Ors-LiF-TTNS,MIT-Fertita-2014,Corinne_2015}.

The entanglement entropy of orbital $i$, also called single-orbital entropy, can be calculated from the eigenvalues of the one-orbital reduced density matrix $\omega_{\alpha;i}$~\cite{legeza_dbss},
\begin{equation}\label{eq:s1}
s_i=-\sum_{\alpha=1}^{4}{\omega_{\alpha;i} \ln \omega_{\alpha;i}}.
\end{equation}
The single-orbital entropy is thus the von Neumann entropy of the reduced density matrix of the orbital of interest whose elements can be calculated from the one- and two-particle reduced density matrices~\cite{Kasia_ijqc}, $\gamma^p_q$ and $\Gamma^{pq}_{rs}$, where for a given wavefunction $|\Psi\rangle$
\begin{equation}
\gamma_q^p = \frac{\langle \Psi| a_p^\dagger a_q | \Psi \rangle}{\langle\Psi|\Psi\rangle},
\end{equation}
and
\begin{equation}
\Gamma_{rs}^{pq} = \frac{\langle \Psi| a_p^\dagger a^\dagger_q a_s a_r | \Psi \rangle}{\langle\Psi|\Psi\rangle},
\end{equation}
or from generalized correlation functions~\cite{Barcza2013,entanglement_bonding_2013}.
The one-orbital reduced density matrix $\rho_i$ is spanned by the basis states of the one-orbital Fock space and is thus a $4\times4$ matrix.
Similarly, the entanglement of two orbitals is quantified by the two-orbital entropy $s_{i,j}$,
\begin{equation}\label{eq:s2ij}
s_{i,j}=-\sum_{\alpha=1}^{16}{\omega_{\alpha;i,j}\ln \omega_{\alpha;i,j}},
\end{equation}
where $\omega_{\alpha;i,j}$ are the eigenvalues of the two-orbital reduced density matrix $\rho_{i,j}$, which is defined in terms of basis states of a two-orbital Fock space (16 possible states in the case of spatial orbitals).
In contrast to $\rho_i$, the matrix elements of $\rho_{i,j}$ can be written in terms of the elements of the 1-, 2-, 3-, and 4-particle reduced density matrices, $\gamma^p_q$, $\Gamma^{pq}_{rs}$, $\Gamma^{pqr}_{stu}$, and $\Gamma^{pqrs}_{tuvw}$, with
\begin{equation}
\Gamma^{pqr}_{stu} = \frac{\langle \Psi| a_p^\dagger a^\dagger_q a_r^\dagger a_u a_t a_s | \Psi \rangle}{\langle\Psi|\Psi\rangle},
\end{equation}
and
\begin{equation}
\Gamma^{pqrs}_{tuvw} = \frac{\langle \Psi| a_p^\dagger a^\dagger_q a_r^\dagger a_s^\dagger a_w a_v a_u a_t | \Psi \rangle}{\langle\Psi|\Psi\rangle},
\end{equation}
Given $s_i$ and $s_{i,j}$, we can quantify the correlations between two orbitals $i$ and $j$ by the orbital-pair mutual information,~\cite{Rissler2006,legeza_dbss,Legeza2006}
\begin{equation}\label{eq:i12}
 I_{i|j}=s_i+s_j-s_{i,j},
\end{equation}
which includes correlaions of both classical and quantum origin.
It is generally accepted that the mutual information measure pairwise correlations.
In this work, we will employ the orbital-pair mutual information as a correlation index to quantify orbital-pair correlations embedded in wavefunctions constructed from electron-pair states.

This work is organized as follows.
In section~\ref{sec:theory}, we briefly summarize how the one- and two-orbital reduced density matrices can be calculated for seniority-zero wavefunctions, that is, wavefunctions restricted to electron-pair states.
Numerical examples are presented in section~\ref{sec:hubbard} for the one-dimensional Hubbard model with periodic boundary conditions and in section~\ref{sec:rings} for hydrogen rings.
Finally, we conclude in section~\ref{sec:conclusions}.

\section{Correlation functions for seniority-zero wavefunctions}\label{sec:theory}
If the electronic wavefunction is a CI-expansion with pair-excited Slater determinants only, that is a seniority-zero wavefunction, $\rho_i$ and $\rho_{i,j}$ have a particular simple form~\cite{Kasia_ijqc}.
Restricting the wavefunction expansion to either doubly-occupied or unoccupied orbitals, $\rho_i$ reduces to a $2\times2$ matrix, while $\rho_{i,j}$ becomes a $4\times4$ matrix.
Furthermore, for seniority-zero wavefunctions, we can use the relations $\gamma_{p}^{p}=\gamma_{\bar{p}}^{\bar{p}}=\Gamma_{p\bar{p}}^{p\bar{p}}$ and $\Gamma_{p\bar{q}}^{p\bar{q}}={}^4\Gamma_{p\bar{p}q\bar{q}}^{p\bar{p}q\bar{q}}$~\cite{Weinhold1967a} so that only the 1- and 2-particle reduced density matrcies are required to determine $\rho_i$ and $\rho_{i,j}$. Specifically, we have~\cite{Kasia_ijqc}
\begin{equation}\label{eq:rho1-aprog}
\rho_i =
\begin{pmatrix}
1-\gamma_{i}^{i} & 0 \\
0 & \gamma_{i}^{i} 
\end{pmatrix}
\end{equation}
for the seniority-zero one-orbital RDM expressed in the basis $\{$\zero,\double$\}$, and
\begin{equation}\label{eq:rho2-aprog}
\rho_{i,j} =
\begin{pmatrix}
1-\gamma_{i}^{i} -\gamma_{j}^{j}+\Gamma_{i\bar{j}}^{i\bar{j}} & 0 & 0 & 0 \\
0 & \gamma_{i}^{i}-\Gamma_{i\bar{j}}^{i\bar{j}} & -\Gamma^{i\bar{i}}_{j\bar{j}} & 0\\
0 & -\Gamma^{j\bar{j}}_{i\bar{i}} & \gamma_{j}^{j}-\Gamma_{j\bar{i}}^{j\bar{i}} & 0\\
0 & 0 & 0 &\Gamma_{i\bar{j}}^{i\bar{j}} 
\end{pmatrix}
\end{equation}
for the seniority-zero two-orbital RDM expressed in the basis  $\{$\zero\,\zero,\zero\,\double,\double\,\zero,\double\,\double$\}$. We should note that, for a seniority-zero wavefunction, the maximum value of $s_i$ is $\ln 2$.

For AP1roG, the response 1- and 2-particle RDMs are used to construct $\rho_i$ and $\rho_{i,j}$ and are defined as
\begin{equation}
    \gamma^p_q = \langle \Phi_0 \vert (1+\Lambda) e^{-\hat{T}_p} \{a_p^\dagger a_q\}e^{\hat{T}_p} \vert \Phi_0 \rangle
\end{equation}
and
\begin{equation}
    \Gamma^{pq}_{rs} = \langle \Phi_0 \vert (1+\Lambda) e^{-\hat{T}_p} \{a_p^\dagger a_q^\dagger a_s a_r\}e^{\hat{T}_p} \vert \Phi_0 \rangle,
\end{equation}
where $\Lambda=\sum_{ia} \lambda_a^i a^\dagger_i a^\dagger_{\bar{i}} a_{\bar{a}} a_a$ is a de-excitation operator and $\hat{T}_p$ is restricted to pair excitations only (cf.~eq.~\eqref{eq:ap1rog}).
Furthermore, due to the special structure of the wavefunction, the only non-zero elements are $\gamma_{p}^{p}$, $\Gamma_{pq}^{pq}$, and $\Gamma_{pp}^{qq}$.
We should note that the response density matrices are not Hermitian and, in general, we have $\Gamma_{pp}^{qq} \neq \Gamma_{qq}^{pp}$.
The deviation from Hermiticity of the response density matrices is an artefact of the truncation of the full cluster operator and disappears if the full cluster operator is taken in the coupled cluster ansatz.
As the AP1roG method uses, however, a truncated cluster operator, we cannot exclude non-symmetric two-particle response density matrices.
Furthermore, if the response density matrices are not symmetric and are thus not $N$-representable, the resulting eigenvalues of $\rho_{p,q}$ might result in negative values for orbital pair $p,q$.
In this work, however, we haven't observed any problems with $N$-representability of the response density matrices if the orbital basis is optimized within the AP1roG method.
Only minor $N$-representability issues have been observed when using canonical Hartree--Fock orbitals in the strong correlation regime with negative eigenvalues of order $10^{-3}$ or much smaller (see also section~\ref{sec:rings}).
Since negative eigenvalues are unphysical, we have discarded them when calculating the correlation functions.

\begin{figure}[htbp]
\centering
\includegraphics[width=0.99\linewidth]{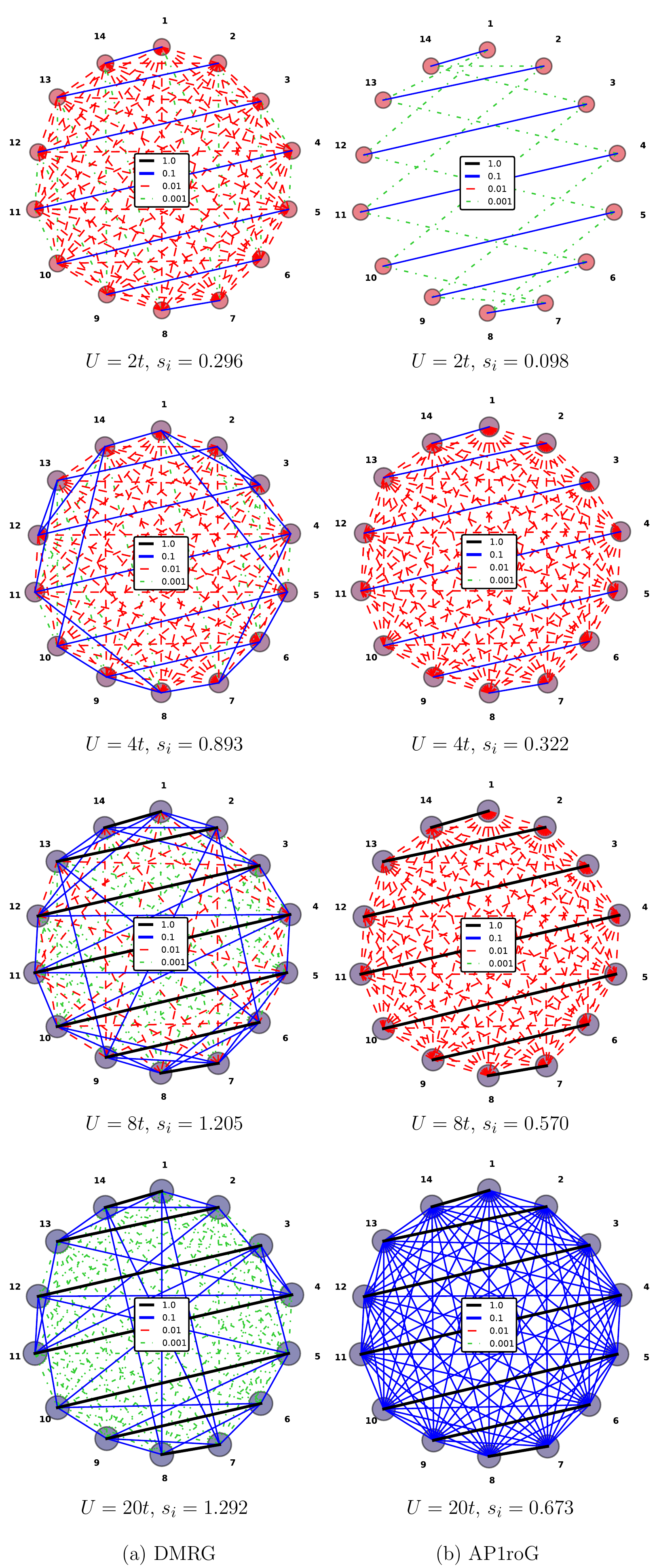}
\caption{(Color online)
Orbital-pair mutual information for the half-filled 1-D Hubbard model with periodic boundary conditions, 14 sites, and different on-site interaction strengths for the optimized AP1roG basis.
The single-orbital entropy is site-independent and given below each figure.
The strength of the orbital-pair correlations for both the (a) DMRG (left panel) and (b) AP1roG (right panel) correlation diagrams are color-coded: black lines indicate strong correlations, while green lines indicate weak correlations.
}
\label{fig:hub1}
\end{figure}

\begin{figure*}[htbp]
\centering
\includegraphics[width=0.9\linewidth]{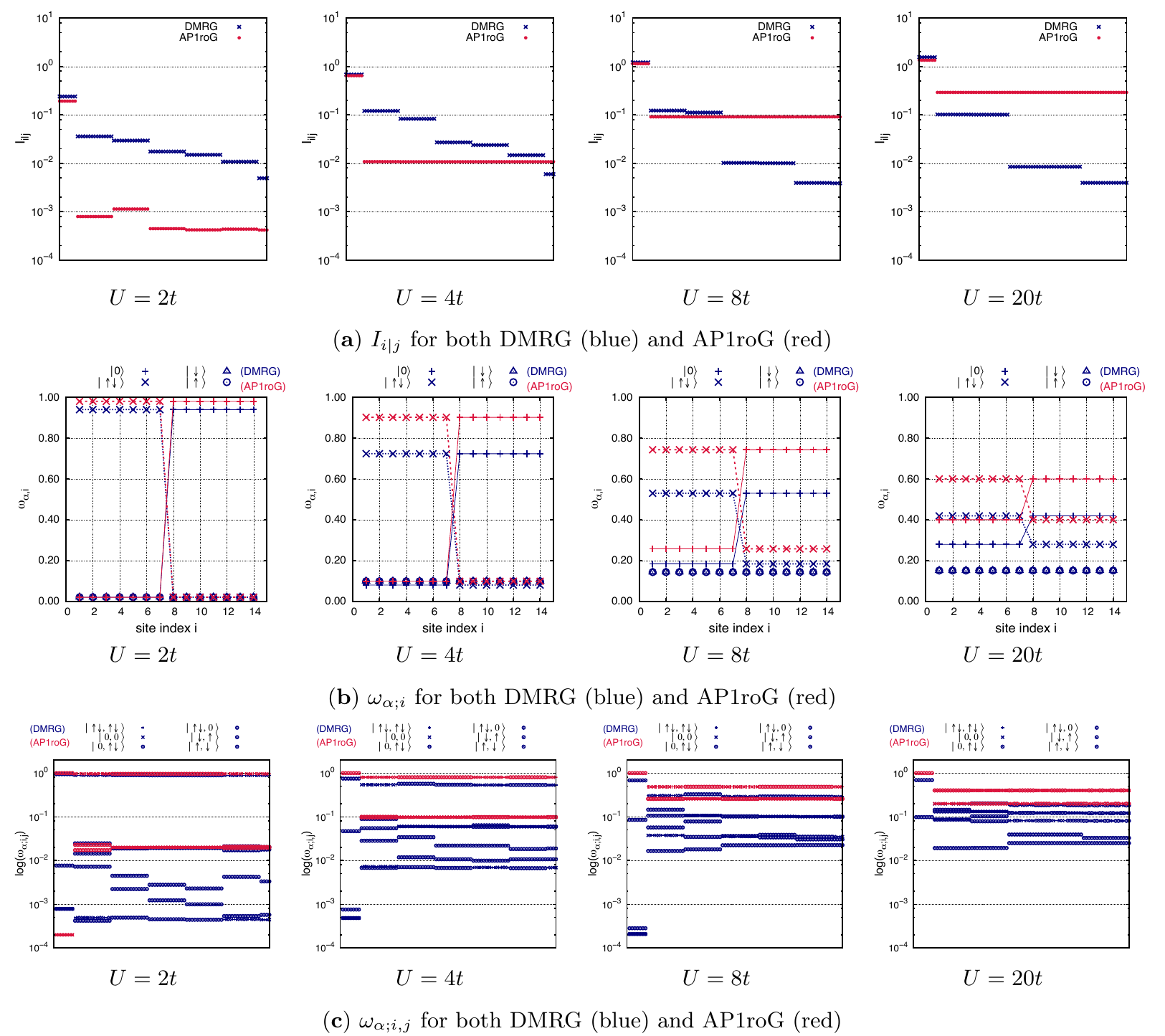}
\caption{(Color online)
Decaying values of the mutual information for the half-filled Hubbard model with 14 sites using the optimized AP1roG orbital basis (a).
$I_{i|j}$ is sorted with respect to the DMRG reference values so that each value of $I_{i|j}$ is shown for the same orbital pair $i$ and $j$ in both DMRG and AP1roG calculations.
Eigenvalues of the (b) one-orbital reduced density matrix and (c) two-orbital reduced density matrix for the half-filled Hubbard model with 14 sites obtained in DMRG and AP1roG calculations using the optimized AP1roG orbital basis.
The eigenvalues of $\rho_{i,j}$ for each pair $i,j$ are ordered as in (a).
Red lines and symbols indicate AP1roG data, while blue lines and symbols mark the corresponding DMRG results.
}
\label{fig:hub2}
\end{figure*}

\section{The half-filled one-dimensional Hubbard Hamiltonian}\label{sec:hubbard}
First, we consider the 1-D Hubbard model Hamiltonian with periodic boundary conditions,
\begin{equation}
\hat{H}_{\rm Hub} = -t\sum_{\substack{j\\\sigma \in \{\uparrow, \downarrow\}}} \left( a_{(j+1)\sigma}^{\dagger}a_{j\sigma} + a_{j\sigma}^{\dagger}a_{(j+1)\sigma} \right ) 
+U\sum_j n_{j\uparrow} n_{j\downarrow},
\end{equation}
where the first term describes nearest-neighbor hopping, while the second term represents the repulsive on-site interaction. 
$n_{j\sigma} = a_{j\sigma}^{\dagger}a_{j\sigma}$ is the local number operator.
It is well-known that the one-dimensional half-filled Hubbard model for $U=0t$ is gapless, where all four local basis states ($|$\zero$\rangle$,$|$\up$\rangle$,$|$\down$\rangle$,$|$\double$\rangle$) have equal weights $\frac{1}{4}$ and hence the site entropy $s_i=\ln(4)$.
For $U>0t$, the charge gap opens and the weight of the unoccupied and doubly-filled basis states decrease.
In the large $U/t\rightarrow \infty$ limit, only the $|$\up$\rangle$ and $|$\down$\rangle$ states have weights of 0.5 with $s_i=\ln(2)$ as the model becomes equivalent to the spin-$\nicefrac{1}{2}$ Heisenberg model and the ground state is an antiferromagnetic state.
Therefore, a wavefunction restricted to electron-pair states ($|$\zero$\rangle$ and $|$\double$\rangle$) cannot properly describe the correlations in both the large $U/t$ limit and, to a smaller extent, for small $U/t$ using the local on-site basis.
To properly model such wavefunctions, we have to change the basis, which allows us to describe correlations of the one-dimensional half-filled Hubbard model with only unoccupied and doubly-filled basis states.
Such a basis can be obtained self-consistently within the AP1roG method as, for instance, described in Refs.~\cite{OO-AP1roG,Tamar-pCC,PS2-AP1roG,AP1roG-JCTC}.
Note that correlation and entanglement measures are basis dependent and thus the one-site(orbital) and two-orbital correlations within the optimized AP1roG basis will differ from those in the local on-site basis.
In order to assess the accuracy of AP1roG in describing orbital-pair correlations of the one-dimensional half-filled Hubbard model, we will perform DMRG calculations using the optimized AP1roG basis.
As an example, we will only investigate the one-dimensional Hubbard model with 14 sites. 
The electronic energies obtained by DMRG and AP1roG as well as additional numerical examples using 30 sites are summarized in the Supporting Information.

\subsection{The Hubbard model in the AP1roG basis}
Figure~\ref{fig:hub1} shows the orbital-pair mutual information and the single-orbital entropy obtained from DMRG (left panel) and AP1roG (right panel), respectively, for different strengths of $U$ for the optimized AP1roG basis.
For all investigated values of $U$, AP1roG can reproduce the most important orbital correlations (\textit{cf.}~the black/blue lines in Figure~\ref{fig:hub1}).
Weaker orbital correlations ($I_{i|j}\leq10^{-2}$) are, however, underestimated for small $U/t$ if the wavefunctions is restricted to the seniority-zero sector.
For increasing repulsive on-site interactions $U\geq 4t$, AP1roG gradually overestimates orbital-pair correlations compared to the DMRG reference distribution (\textit{cf.}~increasing number of red/blue lines).
To emphasize the observed overcorrelation of AP1roG for increasing $U/t$, Figure~\ref{fig:hub2}(a) displays the decaying values of $I_{i|j}$ obtained by DMRG and AP1roG. Each point on the graph is plotted for the same orbital pair $(i,j)$ and all orbital-pair correlations are sorted with respect to the DMRG reference values.
While AP1roG predicts a decaying trend of $I_{i|j}$ qualitatively similar to the DMRG reference curve for $U=2t$, it underestimates a large amount of orbital-pair correlations by more than an order of magnitude.
Increasing $U$ to $4t$ leads to the formation of plateaus in $I_{i|j}$, which become more pronounced the stronger the on-site interaction.
In contrast to DMRG, AP1roG does not feature the multiple characteristic steps in $I_{i|j}$, but rather shows a prolonged plateau of orbital-pair correlations.
This plateau moves upwards to larger values of $I_{i|j}$ when $U$ increases.
While the overestimation of orbital-pair correlations is minor for intermediate on-site repulsion strengths $U\leq4t$, $I_{i|j}$ is overestimated for all orbital pairs $(i,j)$ (and $I_{i|j} < 1$) when $U$ reaches the strong correlation limit.
Thus, restricting the wavefunction to the seniority-zero sector results in an overestimation of the medium-sized and weak orbital-pair correlations.

To elucidate the origins of these discrepancies, we will analyze the eigenvalues and eigenvectors of the one- and two-orbital reduced density matrices $\rho_{i}$ and $\rho_{i,j}$, respectively, obtained from DMRG and AP1roG calculations.
Figure~\ref{fig:hub2}(b) shows the spectrum of $\rho_{i}$ for each site index $i$.
Note that, for AP1roG, $\rho_{i}$ is a $2\times2$ matrix, while DMRG also includes the spin-up and spin-down contributions and is thus represented by a $4\times4$ matrix.
Furthermore, since we have introduced a reference determinant $|\Phi_0\rangle$ that differes between occupied ($i\in \{1,2,\ldots,7\}$) and virtual ($i\in \{8,9,\ldots,14\}$) orbitals, the corresponding site/orbital entropies are not equivalent.
This is also evident from eq.~\eqref{eq:rho1-aprog}, which contains the natural occupation numbers $\gamma_i^i$ with $\gamma_i^i\approx2$ for occupied orbitals and $\gamma_i^i\approx0$ for virtual orbitals, respectively.
For small $U/t$, the eigenvalues $\omega_{\alpha,i}$ are either close to one or close to zero and the spectrum of $\rho_i$ obtained by AP1roG is qualitatively similar to the DMRG reference.
However, the differences in $\omega_{\alpha,i}$ accumulate when reaching the strong correlation limit.
Specifically, increasing the on-site repulsion, changes the contributions corresponding to the basis states $|$\zero$\rangle$ and $|$\double$\rangle$, while the eigenvalues of the singly-occupied states ($|$\up$\rangle$ and $|$\down$\rangle$) gradually increase.
For large $U/t$, the contributions of $|$\up$\rangle$ and $|$\down$\rangle$ to $s_i$ are significant and similar in magnitude to $|$\zero$\rangle$ and $|$\double$\rangle$.
Thus, in the strong correlation limit, the contributions of $|$\up$\rangle$ and $|$\down$\rangle$ are important and the wavefunction cannot be accurately described by the seniority-zero sector alone.

Similar conclusions can be drawn from the spectrum of $\rho_{i,j}$.
Figure~\ref{fig:hub2}(c) shows the eigenvalues of $\rho_{i,j}$ for each orbital pair $i,j$. The eigenvalues for each pair $i,j$ are ordered as in Figure~\ref{fig:hub2}(a).
Note that blocks spanned by states that preserve the particle number $n$ and the $s_z$ quantum number are decoupled.
Thus, states spanned by $|$\zero,\zero$\rangle$ and $|$\double,\double$\rangle$ represent two (uncoupled) eigenvectors with $(n,s_z)=(0,0)$ and $(n,s_z)=(4,0)$, respectively, while the states $|$\zero,\double$\rangle$ and $|$\double,\zero$\rangle$ couple to the states $|$\up,\down$\rangle$ and $|$\down,\up$\rangle$, which corresponds to the subblock $(n,s_z)=(2,0)$.
The latter basis vectors ($|$\up,\down$\rangle$ and $|$\down,\up$\rangle$) always have zero contributions if the wavefunction is restricted to electron pair states, and hence, only the $|$\zero,\double$\rangle$ and $|$\double,\zero$\rangle$ sectors are coupled.
Due to the coupling between $|$\up,\down$\rangle$, $|$\down,\up$\rangle$, $|$\zero,\double$\rangle$, and $|$\double,\zero$\rangle$, we cannot assign an eigenvalue to a specific local state and the corresponding eigenvalues are marked by the same symbol in Figure~\ref{fig:hub2}(c).
We should note, however, that the coupling between the doubly occupied/unoccupied and singly-occupied states is, in general, small, becomes, however, non-negligible for large values of $U/t$.
If not mentioned otherwise, we will distinguish between eigenvalues corresponding to eigenvectors with dominant contributions from the $|$\down,\up$\rangle$ and $|$\up,\down$\rangle$ states (referred to as singly-occupied states in the following) and from the $|$\zero,\double$\rangle$ and $|$\double,\zero$\rangle$ states (identified as doubly-occupied/unoccupied states).
Moreover, we will restrict our analysis of $\rho_{i,j}$ to the three sub-blocks $(n,s_z)=\{(0,0),(2,0),(4,0)\}$ spanned by $|$\up,\down$\rangle$, $|$\down,\up$\rangle$, $|$\zero,\double$\rangle$, $|$\double,\zero$\rangle$, $|$\zero,\zero$\rangle$, and $|$\double,\double$\rangle$ as these sub-blocks (excluding the singly-occupied states) are non-zero for the AP1roG wavefunction.
The complete eigenvalue spectrum obtained by DMRG calculations is summarized in the Supporting Information.

As observed for $\rho_{i}$, AP1roG reproduces the largest eigenvalues ($\omega_{\alpha;i,j}>0.01$) for small $U/t$.
Thus, the dominant part of the spectrum of $\rho_{i,j}$ can be described by electron-pair states, while contributions from singly-occupied states are approximately one order of magnitude smaller ($\omega_{\alpha;i,j}\ll0.01$).
If $U/t$ increases, the differences in $\omega_{\alpha;i,j}$ between AP1roG and DMRG increase.
In general, the dominant eigenvalues of $\rho_{i,j}$ are overestimated in AP1roG compared to the DMRG reference.
Simultaneously, $\omega_{\alpha;i,j}$ attributed to the singly-occupied states increase considerably.
In the strong correlation limit, such singly-occupied states become important (their contributions to the eigenvalue spectrum increases by approximately one order of magnitude), especially for the description of weak orbital-pair correlations ($I_{i|j}\le0.001$).
If these singly-occupied states are excluded in the wavefunction expansion (as in wavefunctions built from electron-pair states), the spectrum of $\rho_{i,j}$ cannot be properly described (note the plateau in $\omega_{\alpha;i,j}$ for AP1roG in Figure~\ref{fig:hub2}(a)).
Thus, the two-orbital entropy is underestimated, which, in turn, overvalues the orbital-pair mutual information for small $I_{i|j}\le0.001$ (cf.~Figures~\ref{fig:hub1} and \ref{fig:hub2}(a)).
We should emphasize that the largest orbital-pair correlations are accurately reproduced by AP1roG for large on-site interaction strenghts.
For these orbital-pair correlations, however, only the eigenvalues of $\rho_{i}$ contribute to $I_{i|j}$ as $\omega_{\alpha;i,j}\approx1.0$ (cf.~Figure~\ref{fig:hub2}).

\begin{table}[tbp]
\caption{Selected eigenvalues and eigenvectors of $\rho_{i,j}$ for selected orbital-pairs of the one-dimensional Hubbard model with 14 sites and different on-site interaction strengths. Only the largest of the four eigenvectors of the $(n,s_z)=(2,0)$ sub-block are shown.}
\label{tab:hub1}
{
\begin{center}
\begin{tabular}{ccc| rr rr}
\hline \hline
& & & \multicolumn{4}{c}{$(n,s_z)=(2,0)$} \\ 
$(i,j)$ & $U/t$& $\omega_{\alpha;i,j}$ & $|$\zero,\double$\rangle$ &  
                                $|$\double,\zero$\rangle$ &
                                $|$\down,\up$\rangle$ &
                                $|$\up,\down$\rangle$ \\ \hline \hline
\multirow{4}{*}{1,2} &2 & 0.025 &  0.690&  0.690&$-$0.155&   0.155 \\
  &4 &0.064 &   0.683&   0.683&$-$0.184&  0.184 \\
  &8 &0.146 &$-$0.571&$-$0.571&$-$0.418&  0.418 \\
  &20&0.145 &   0.636&   0.636&   0.308&$-$0.308 \\\hline
\multirow{4}{*}{1,13} &2 &0.005&$-$0.052 &   0.040 &$-$0.706 &   0.706\\
 &4 &0.022 &$-$0.141&   0.088 &$-$0.697 &   0.697\\
 &8 &0.078 &   0.301&$-$0.219 &$-$0.656 &   0.656\\
 &20&0.103 &$-$0.635&   0.169 &   0.533 &$-$0.533\\
\hline \hline
\end{tabular}
\end{center}
}
\end{table}

Finally, we should note on the coupling between doubly-occupied/empty and singly-occupied states.
While for all $U/t$, the $|$\up,\down$\rangle$/$|$\down,\up$\rangle$ and $|$\zero,\double$\rangle$/$|$\double,\zero$\rangle$ states are uncoupled for the largest orbital-pair correlations (for instance indices 1/14, 2/13, etc. in Figure~\ref{fig:hub1}), for intermediate and weak correlations the coupling increases with increasing $U/t$ (see Table~\ref{tab:hub1}).

\section{Dissociation of hydrogen rings}\label{sec:rings}
\begin{figure}[!]
\centering
\includegraphics[width=0.9\linewidth]{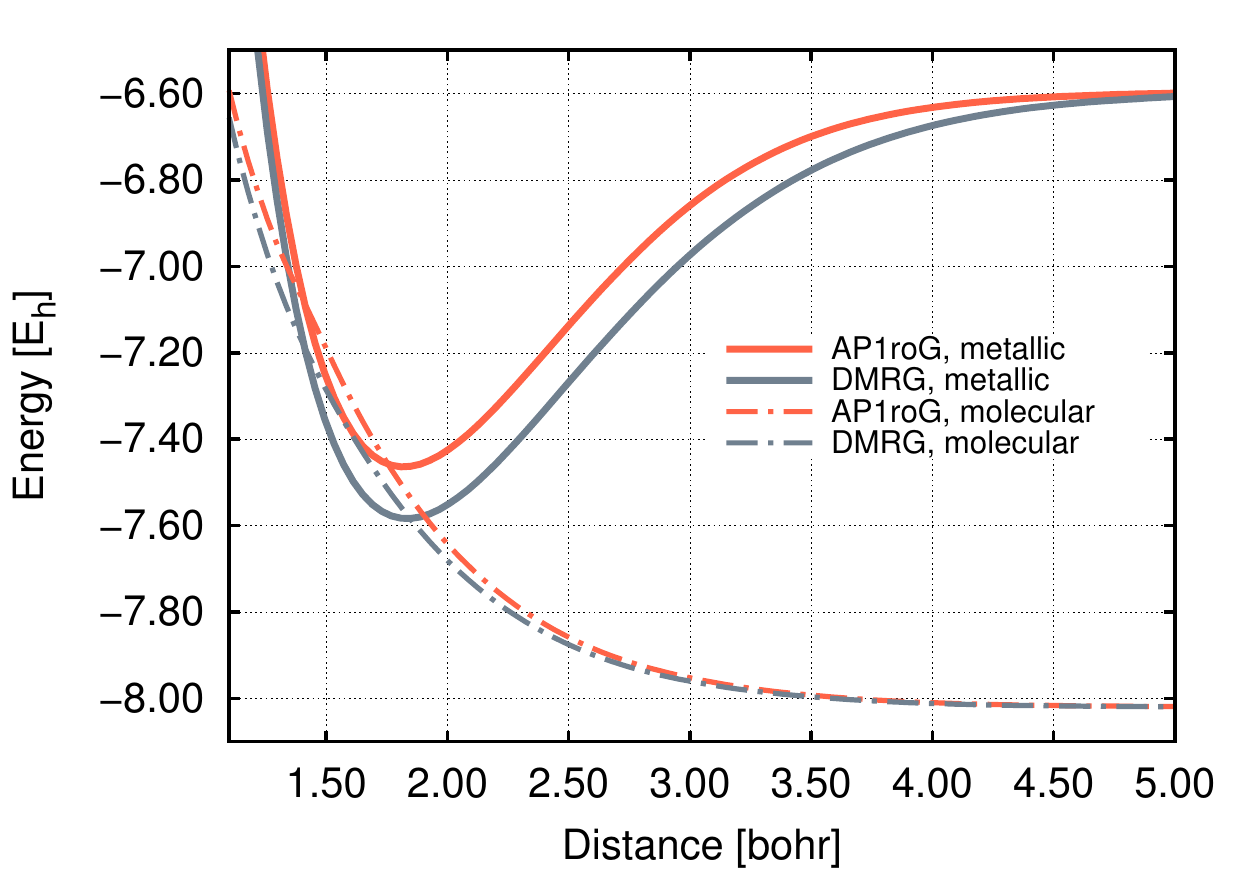}
\caption{(Color online)
DMRG and AP1roG potential energy surfaces for the dissociation of molecular and metallic H$_{14}$ using the STO-6G basis set.
}
\label{fig:pes}
\end{figure}

\begin{figure*}[!]
\centering
\includegraphics[width=0.99\linewidth]{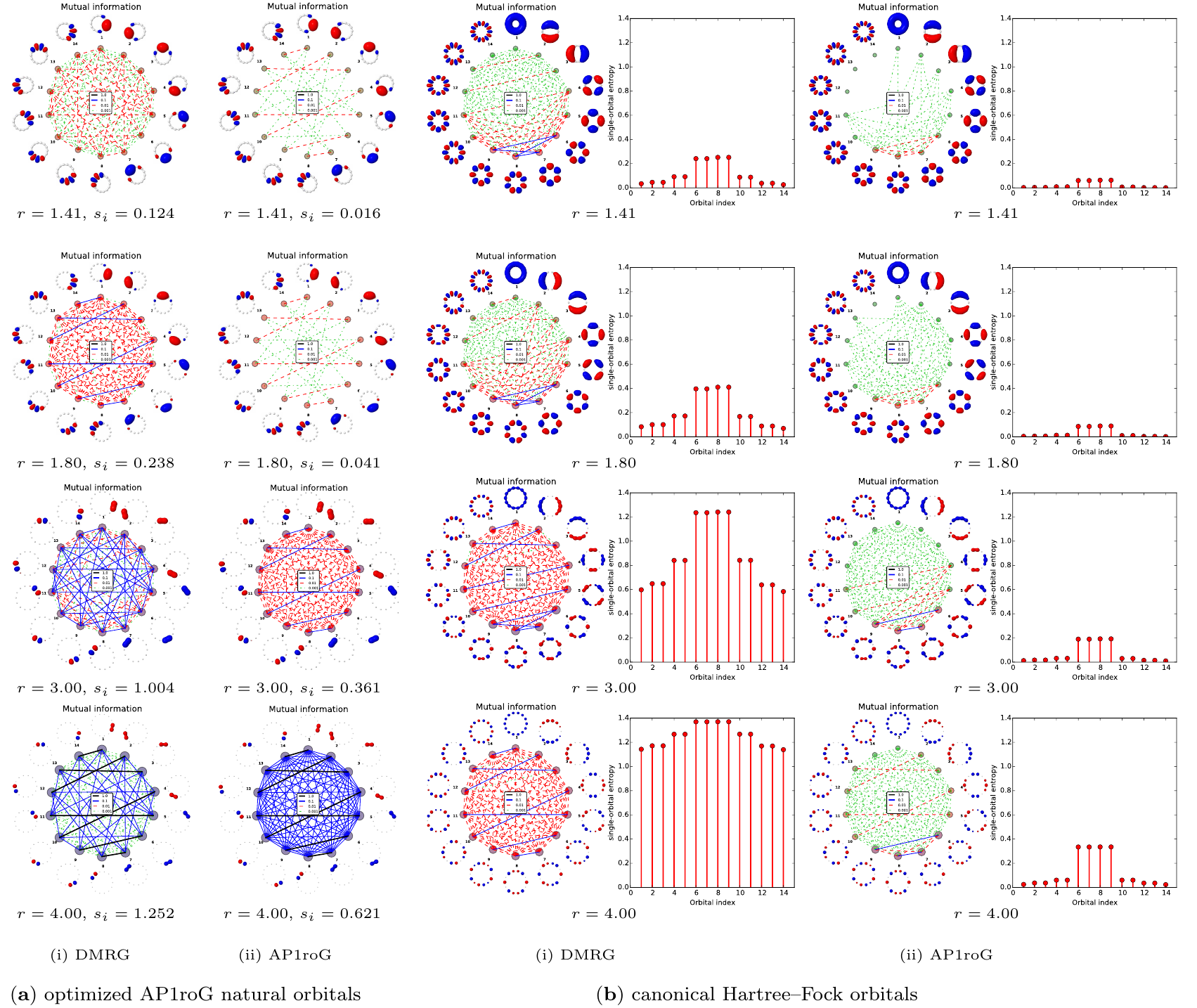}
\caption{(Color online)
Orbital-pair mutual information and single-orbital entropy for the metallic H$_{14}$ ring.
Both the (i) DMRG (left panel) and (ii) AP1roG (right panel) correlation diagrams are obtained for the (a) optimized AP1roG orbital basis and (b) canonical Hartree--Fock orbital basis.
The strength of the orbital-pair correlations are color-coded: black lines indicate strong correlations, while green lines indicate weak correlations.
The orbitals corresponding to each index $i$ are also shown in the mutual information plot.
}
\label{fig:met1}
\end{figure*}

\begin{figure*}[!]
\centering
\includegraphics[width=0.9\linewidth]{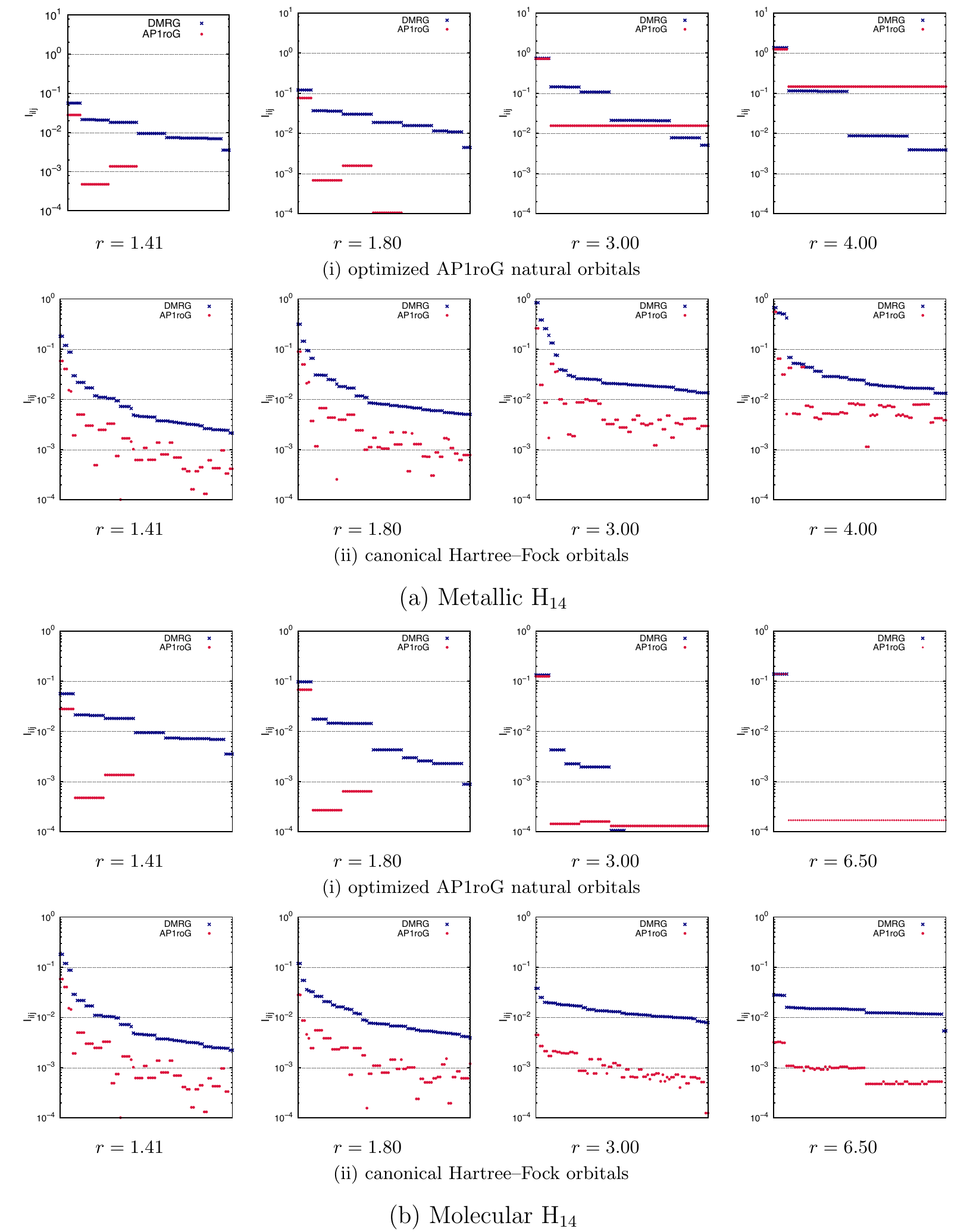}
\caption{(Color online)
Decaying values of the mutual information for the (a) metallic H$_{14}$ ring and (b) molecular H$_{14}$ ring.
For each hydrogen ring, we have used (i) optimized AP1roG natural orbitals and (ii) canonical Hartree--Fock orbitals.
$I_{i|j}$ is sorted with respect to the DMRG reference values so that each value of $I_{i|j}$ is shown for the same orbital pair $i$ and $j$ in both DMRG and AP1roG calculations.
}
\label{fig:i12-ring}
\end{figure*}

Our next numerical example is the dissociation of hydrogen rings.
Specifically, we investigate the symmetric and asymmetric stretching of the H$_{14}$ molecule.
The symmetrically stretched hydrogen ring will be refered to as metallic H$_{14}$, while the asymmetrically stretched H$_{14}$ ring will be indicated as the molecular hydrogen ring as the dissociation process will result in separated hydrogen molecules~\cite{Seel1982,Liegener1985,Anndre1988,Wright1992}.
Furthermore, in molecular H$_{14}$, the distance between the hydrogen atoms of the separated hydrogen molecules was kept fixed at 1.41 bohr in accordance with Ref.~\cite{Seel1982}.
For the symmetric and asymmetirc dissociation of H$_{14}$, the quantum system is described by the non-relativistic quantum chemical Hamiltonian, which reads in its second quantized form 
\begin{equation}\label{eqn:Chem}
\hat{H} = \sum_{pq,\sigma} h_{pq}  a^\dagger_{p\sigma} a_{q\sigma} + \frac{1}{2} \sum_{pqrs,\sigma\tau} \langle pq|rs \rangle a^\dagger_{p\sigma} a^\dagger_{q\tau} a_{s\tau} a_{r\sigma}+H_{\rm nuc},
\end{equation}
where the first term contains both the kinetic energy and nuclear--electron attraction, the second term represents the repulsive electron-electron interaction, and the third term is the nuclear--nuclear repulsion energy, respectively.  
The indices $p$, $q$, $r$, and $s$ run over all one-particle basis functions.
By changing the distances between H atoms and H$_2$ molecules in hydrogen rings, we simulate the one-dimensional Hubbard (metallic H$_{14}$) and dimerized Hubbard model (molecular H$_{14}$) as a function of the on-site interaction $U$ using an \textit{ab initio} treatment.
We should note that we will again modify the on-site localized basis, that is, STO-6G, self-consistently.
As for the Hubbard model, the final basis states used in our numerical calculations thus do not correspond to an on-site localized basis.
To emphasize the differences, the optimized basis states are shown in the orbital-pair correlation graphs (see below).

The potential energy surfaces for the dissociation of metallic and molecular H$_{14}$ using the STO-6G basis set are shown in Figure~\ref{fig:pes}.
While metallic H$_{14}$ has an energy minimum around $r_{\rm H-H}=1.8$ bohr, the total energy of molecular H$_{14}$ gradually decreases for increasing inter-molecular H--H distances. Both curves cross at $r_{\rm H-H}=1.41$ bohr and around the energy minimum of metallic H$_{14}$ located at approximately $r_{\rm H-H}=1.8$ bohr.
Note that for $r_{\rm H-H}=1.41$, the molecular structures of the metallic and molecular H$_{14}$ rings are identical and correspond to a ring of equidistant hydrogen atoms.
For both hydrogen rings, the potential energy surfaces predicted by AP1roG agree well with the DMRG reference curves.
Larger deviations can be observed for H--H distances of approximately $r_{\rm H-H}=2.00$ bohr for both metallic and molecular H$_{14}$, with the latter deviating less from the DMRG reference potential energy surface (up to 0.10 $E_h$ for molecular H$_{14}$ compared to 0.14 $E_h$ for metallic H$_{14}$).
These differences can be associated with electron correlation effects that cannot be described by electron-pair states only.
Furthermore, in the vicinity of dissociation, the differences between the predicted AP1roG energy curve and the DMRG reference curve become negligible ($\Delta E\ll0.01\,E_h$).
Note that for the molecular hydrogen ring, the dissociation limit corresponds to separated hyrdrogen molecules for which AP1roG is exact (as for all two-electron systems).
The total electronic energies obtained by DMRG and AP1roG are summarized in the Supporting Information.

\begin{figure*}[!]
\centering
\includegraphics[width=0.9\linewidth]{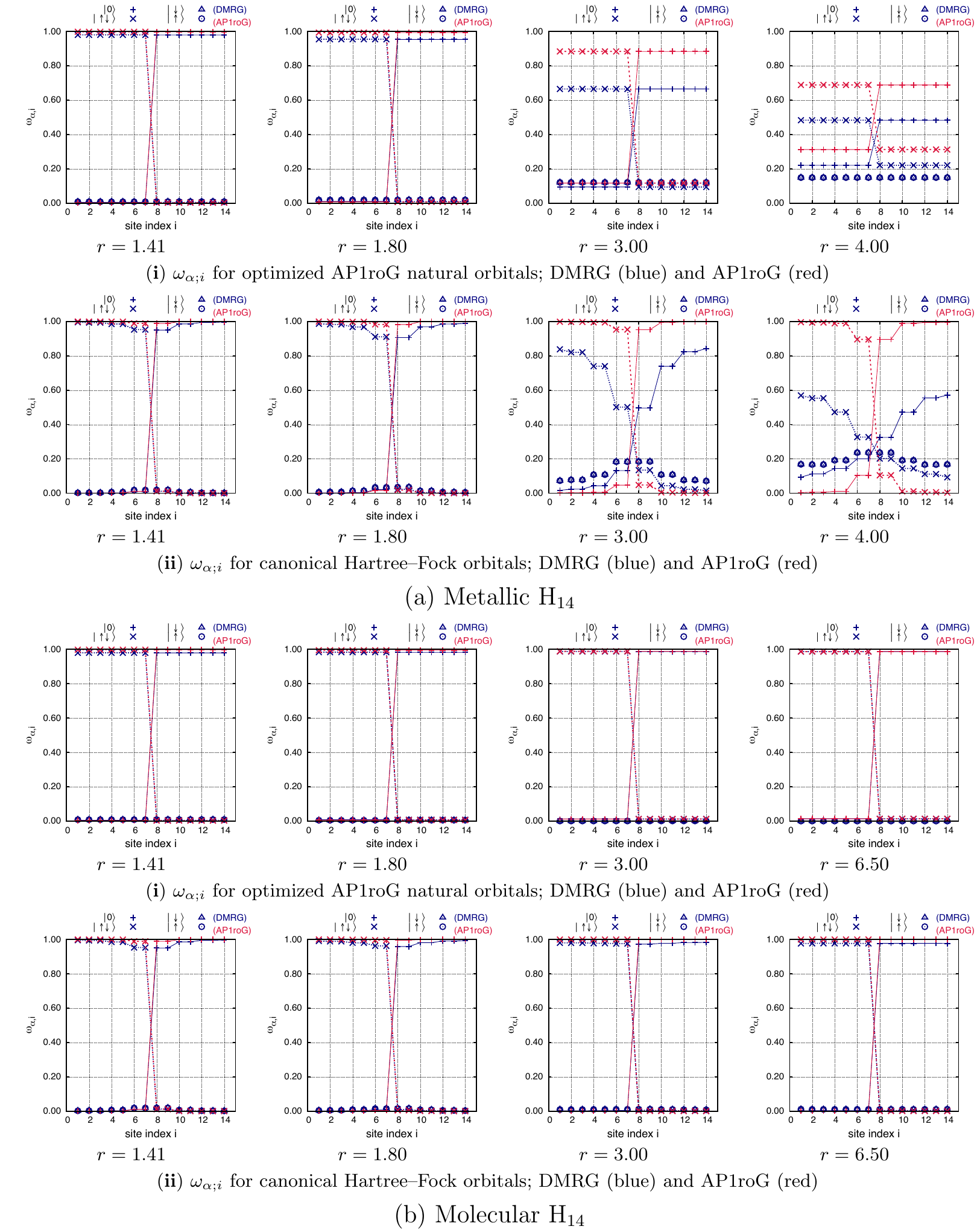}
\caption{(Color online)
Eigenvalues of the one-orbital reduced density matrix for the (a) metallic H$_{14}$ ring and (b) molecular H$_{14}$ ring.
For each hydrogen ring, we have used (i) optimized AP1roG natural orbitals and (ii) canonical Hartree--Fock orbitals.
The orbitals are numbered as in Figure~\ref{fig:met1}.
Red lines and symbols indicate AP1roG data, while blue lines and symbols mark the corresponding DMRG results.
}
\label{fig:rho1-ring}
\end{figure*}

\begin{figure*}[!]
\centering
\includegraphics[width=0.9\linewidth]{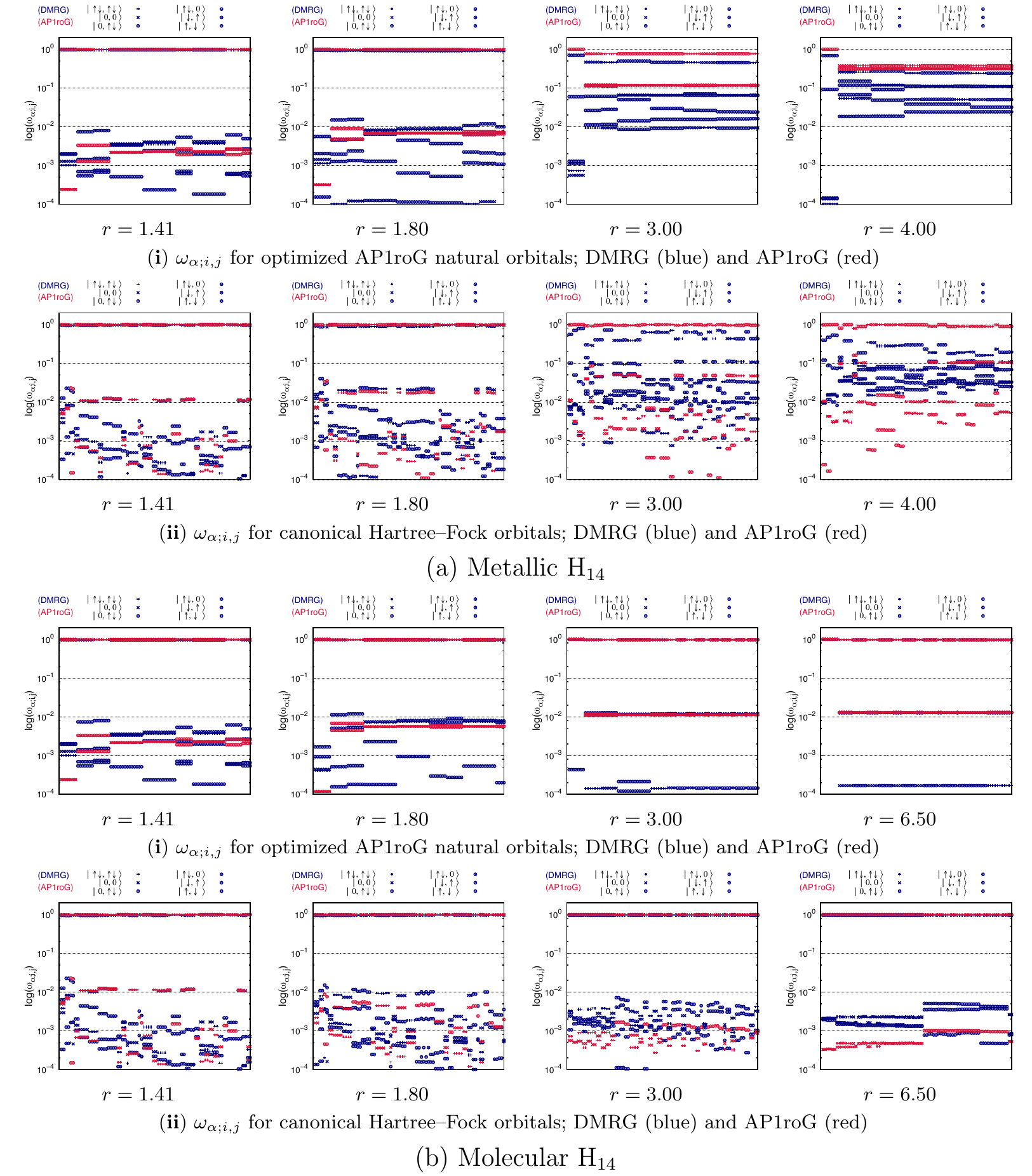}
\caption{(Color online)
Eigenvalues of the two-orbital reduced density matrix for the (a) metallic H$_{14}$ ring and (b) molecular H$_{14}$ ring.
For each hydrogen ring, we have used (i) optimized AP1roG natural orbitals and (ii) canonical Hartree--Fock orbitals.
The eigenvalues are plotted for each orbital pair $(i,j)$ and sorted as the corresponding decaying values of $I_{i|j}$ in Figure~\ref{fig:i12-ring}.
Red lines and symbols indicate AP1roG data, while blue lines and symbols mark the corresponding DMRG results.
}
\label{fig:rho2-ring}
\end{figure*}

\begin{figure*}[!]
\centering
\includegraphics[width=0.99\linewidth]{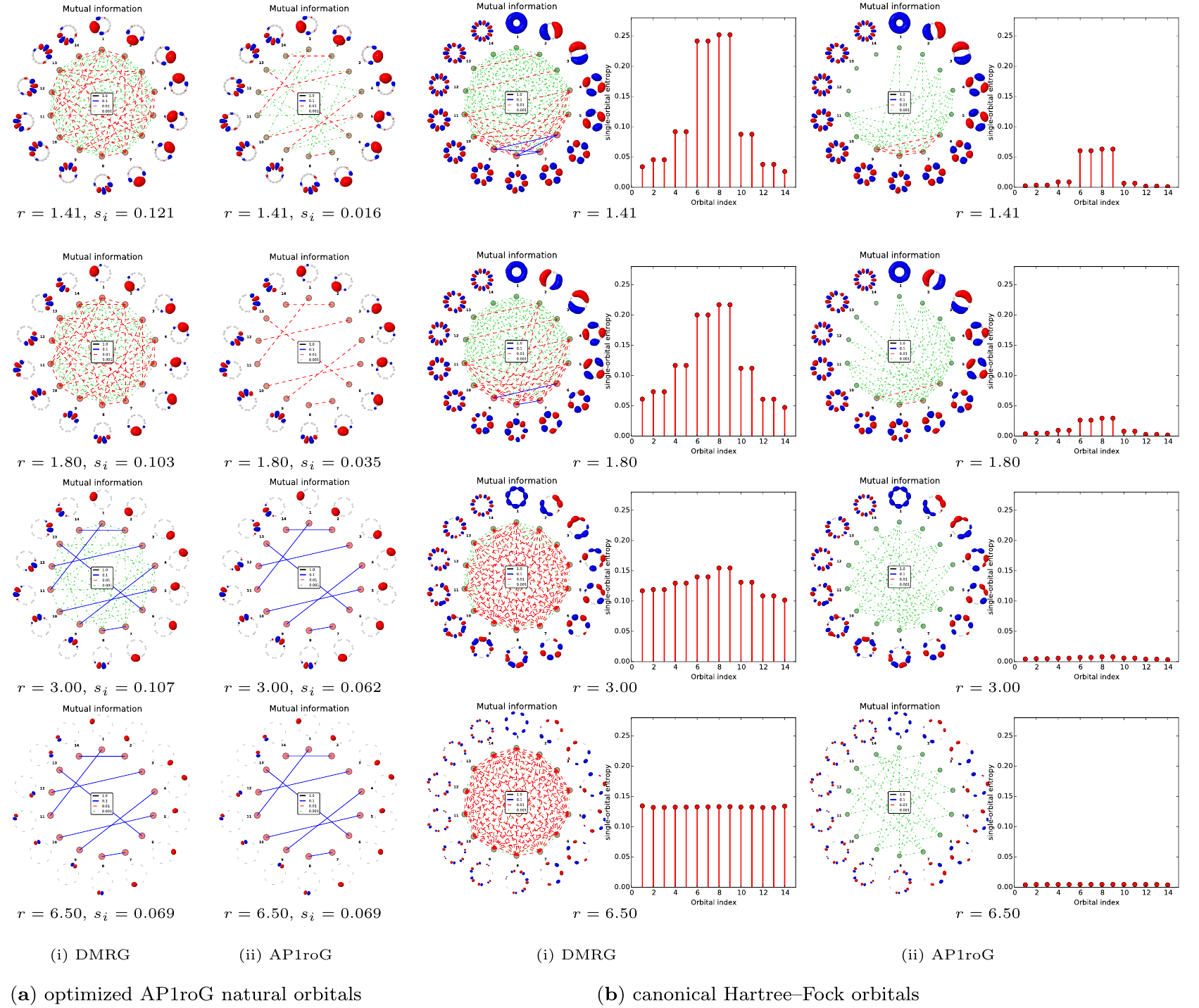}
\caption{(Color online)
Orbital-pair mutual information and single-orbital entropy for the molecular H$_{14}$ ring.
Both the (i) DMRG (left panel) and (ii) AP1roG (right panel) correlation diagrams are obtained for the (a) optimized AP1roG orbital basis and (b) canonical Hartree--Fock orbital basis.
The strength of the orbital-pair correlations are color-coded: black lines indicate strong correlations, while green lines indicate weak correlations.
The orbitals corresponding to each index $i$ are also shown in the mutual information plot.
}
\label{fig:mol1}
\end{figure*}

\subsection{Metallic H$_{14}$}
Figure~\ref{fig:met1} shows the single-orbital entropy and orbital-pair mutual information obtained by AP1roG and DMRG for metallic H$_{14}$ in the optimized AP1roG basis at four characteristic points of the potential energy surface: the squeezed hydrogen ring ($r_{\rm H-H}=1.41$ bohr), around the equilibrium ($r_{\rm H-H}=1.80$ bohr), for a stretched hydrogen ring ($r_{\rm H-H}=3.00$ bohr), and in the vicinity of dissociation ($r_{\rm H-H}=4.00$ bohr).
For short interatomic H--H distances ($r_{\rm H-H}\leq1.80$ bohr), AP1roG (right panel of Figure~\ref{fig:met1}(a)) reproduces the most important orbital-pair correlations between each bonding and antibonding pair of the $\sigma$/$\sigma^*$-orbitals.
However, AP1roG misses a large fraction of the weaker orbital-pair correlations ($I_{i|j}\leq10^{-2}$).
If the hydrogen atoms are pulled further apart ($r_{\rm H-H}=3.00$ bohr), the correlations between the bonding and antibonding $\sigma$/$\sigma^*$-orbitals increase.
Simultaneously, all remaining orbital-pair correlations as predicted by AP1roG accumulate and approach the DMRG reference distribution for all $I_{i|j}\geq 10^{-2}$.
However, weaker orbital-pair correlations (for $I_{i|j}<10^{-2}$) are overestimated compared to the DMRG reference.
The observed overcorrelation  predicted by AP1roG further increases in the vicinity of dissociation ($r_{\rm H-H}\geq4.00$ bohr).
To emphasize the differences in $I_{i|j}$ between DMRG and AP1roG, Figure~\ref{fig:i12-ring}(a-i) shows the decay of $I_{i|j}$ for different interatomic distances sorted with respect to the DMRG reference distribution.
As discussed above, AP1roG cannot describe a large part of the weaker orbital-pair correlations close to the equilibrium structure, while it overestimates weak orbital-pair correlations for a stretched hydrogen ring.
Specifically, AP1roG predicts a prolonged plateau of orbital-pair correlations, in contrast to the stepped decay of $I_{i|j}$ obtained in DMRG calculations.
Note that the orbital-pair correlation diagrams and the decay of the mutual information for metallic H$_{14}$ are qualitatively similar to the one-dimensional Hubbard model using periodic boundary conditions.

Figure~\ref{fig:rho1-ring}(a) shows the eigenvalues of $\rho_{i}$ for each orbital $i$ in metallic H$_{14}$.
As observed in the one-dimensional Hubbard model, AP1roG predicts an eigenvalue spectrum that is qualitative similar to the DMRG refence distribution.
However, for increasing inter-atomic distances ($r_{\rm H-H}\ge3.00$ bohr), the spectrum $\omega_{\alpha;i}$ predicted by AP1roG considerably differs from the DMRG reference.
Specifically, the eigenvalues corresponding to the singly-occuppied states increase by more than one order of magnitude when going from the equilibrium distance to the vicinity of dissocation.
Thus, for large inter-atomic H--H distances, open-shell configurations become important and have to be included in the wavefunctione expanion to reproduce an accureate spectrum of $\rho_{i}$.
The eigenvalues of $\rho_{i,j}$ for the metallic H$_{14}$ ring are plotted in Figure~\ref{fig:rho2-ring}(a).
The eigenvalues are sorted with respect to the magnitude of the mutual information $I_{i|j}$ shown in Figure~\ref{fig:i12-ring}(a).
Similar to the half-filled Hubbard model for small $U/t$, the dominant part of the eigenvalue spectrum ($\omega_{\alpha;i,j}\approx0.01$ or larger) predicted by AP1roG agrees well with the DMRG reference.
In contrast, the eigenvalues corresponding to open-shell configurations are approximately one order of magnitude smaller than those corresponding to the doubly-occupied/unoccupied states ($|$\double,\double$\rangle$, etc.).
However, large differences in $\omega_{\alpha;i,j}$ between AP1roG and DMRG can be found for stretched hydrogen rings.
In the vicinity of dissociation ($r_{\rm H-H}\ge4.00$ bohr), the AP1roG model overestimates all eigenvalues of $\rho_{i,j}$.
Most importantly, the eigenvalues corresponding to singly-occupied states gradually increase when the hydrogen atoms are pulled apart approaching the $\omega_{\alpha;i,j}$ attributed to the doubly-occupied/unoccupied states.
Although differences in total energies decrease, restricting the wavefunction to electron-pair states is insufficient to accurately model the electronic wavefunction for stretched metallic H$_{14}$ and open-shell configurations have to be included in the wavefunction model to reproduce the eigenvalue spectrum of both $\rho_{i}$ and $\rho_{i,j}$.

\subsection{Molecular H$_{14}$}
Similar observation can be made for the dissociation pathway of the molecular hydrogen ring.
The orbital-pair mutual information and the single-orbital entropy for molecular H$_{14}$ are shown in Figure~\ref{fig:mol1}.
For short H--H distances ($r_{\rm H-H}\leq 1.80$ bohr), AP1roG misses a substantial part of the weaker orbital-pair correlations, while the essential correlations between the $\sigma$- and $\sigma^*$-orbitals can be captured by electron-pair states only.
The differences in $I_{i|j}$ between DMRG and AP1roG diminish when the hydrogen molecules are pulled apart.
Furthermore, for increasing inter-molecular H--H distances in the molecular H$_{14}$ ring, the orbital-pair correlations between the $\sigma$- and $\sigma^*$-orbitals localized on each hydrogen molecule gradually increase, while all remaining correlations gradually decrease (see also Figure~\ref{fig:mol1}(a)) for both AP1roG and DMRG.
In the vicinity of dissociation ($r_{\rm H-H}\geq3.00$ bohr), AP1roG slightly overestimates the weakest orbital-pair correlations.
Yet, this overcorrelation is only minor and orders of magnitude smaller than for the metallic hydrogen ring.

Figure~\ref{fig:rho1-ring}(b-i) shows the eigenvalues of $\rho_{i}$ for each orbital $i$ in molecular H$_{14}$.
For increasing inter-molecular H--H distances, the spectrum of $\rho_{i}$ determined by AP1roG approaches the DMRG reference spectrum.
Specifically in the vicinity of separated hydrogen molecules ($r\rightarrow \infty$), the eigenvalues corresponding to the singly-occupied states ($|$\up$\rangle$ and $|$\down$\rangle$) approach zero and the electronic wavefunction can be exactly represented using electron-pair states.
Similar observation can be made for the eigenvalue spectrum of $\rho_{i,j}$ shown in Figure~\ref{fig:rho2-ring}(b-i).
For small inter-molecular distances, singly-occupied states, like, $|$\up,\down$\rangle$, $|$\down,\up$\rangle$ etc., have non-zero $\omega_{\alpha;i,j}$ in the order of 0.001.
Nontheless, AP1roG represents a good approximation for large eigenvalues $\omega_{\alpha;i,j}$.
For stretched molecular H$_{14}$ rings ($r_{\rm H-H}>1.8$ bohr), the eigenvalue spectrum of $\rho_{i,j}$ is dominated by contributions from the doubly-occupied states, which are accurately described using electron-pair states only.
Note that if we approach the regime of (almost) separated hydrogen molecules ($r_{\rm H-H}\approx 6.50$ bohr), the contributions of singly-occupied states to the spectrum of $\rho_{i,j}$ approach zero.
Due to eivenvalues of order $10^{-4}$ for $r_{\rm H-H}=6.50$ bohr, which are zero for seniority-zero wavefunctions, the resulting orbital-pair mutual information exceeds the DMRG reference distribution as the corresponding terms in the two-orbital entropy $s_{i,j}$ vanish (see eqs.~\eqref{eq:s2ij} and \eqref{eq:i12}).

\subsection{The influence of orbital optimization on orbital-pair correlations in H$_{14}$}
Finally, we will focus our discussion on how the orbital optimization affects orbital-pair correlations.
For that purpose we will compare the orbital correlation and entanglement diagrams obtained by DMRG and AP1roG in the canonical Hartree--Fock basis to those calculated in the natural orbital basis optimized within the AP1roG model and the STO-6G basis set.
Figure~\ref{fig:met1}(ii) shows the orbital-pair mutual information and single-orbital entropy for the metallic H$_{14}$ ring obtained in the canonical Hartree--Fock basis for 4 specific points along the dissociation pathway.
For all investigated points, AP1roG misses a substantial amount of (static and dynamic) orbital-pair correlations and orbital entanglement.
However, within the canonical Hartree--Fock basis, AP1roG captures a larger fraction of the weaker orbital-pair correlations (with $I_{i|j}<10^{-2}$) than using the optimized natural AP1roG orbitals (see Figure~\ref{fig:met1}(i)).
Furthermore, all orbital-pair correlations predicted by AP1roG are smaller than the DMRG reference values, which indicates that overcorrelation does not occur when canonical Hartree--Fock orbitals are used to construct the geminals.
These observations are confirmed by the decay of the orbital-pair mutual information displayed in Figure~\ref{fig:i12-ring}(a-ii).
For increasing interatomic H--H distances, we always have $I_{i|j}({\rm AP1roG})<I_{i|j}({\rm DMRG})$.
Despite underestimating a large fraction of the orbital-pair correlations, the decay of $I_{i|j}({\rm AP1roG})$ qualitatively agrees with the DMRG reference distribution for all investigated points along the dissociation pathway.
This is not the case if the orbitals are optimized within the AP1roG method where AP1roG does not predict a stepped decay of $I_{i|j}$ as obtained in DMRG calculations (\textit{cf.}~Figures~\ref{fig:i12-ring}(a-i) and \ref{fig:i12-ring}(a-ii)).

We should note that we have observed $N$-representability problems with the AP1roG response two-particle density matrices for H--H distances $r_{\rm H-H}\ge3.00$ bohr.
This results in negative eingenvalues of $\rho_{i,j}$ for, for instance, orbital pairs $(5,8)$ and $(6,7)$ which slightly increases the corresponding orbital-pair mutual information (as those terms are not subtracted in eq.~\eqref{eq:i12}).
However, this does not significantly influence our conclusions because AP1roG, in general, underestimates orbital-pair correlations within the canonical Hartree--Fock basis.

Figure~\ref{fig:rho1-ring}(a-ii) shows the eigenvalue spectrum of $\rho_{i}$ for each canonical orbital $i$.
For all investigated points of the dissociation pathway, AP1roG predicts eigenvalues $\omega_{\alpha;i}$ that show similar characteristics as the DMRG reference eigenvalues.
However, for stretched hydrogen rings, the differences between AP1roG and DMRG increase and AP1roG cannot reproduce the stepped trend in $\omega_{\alpha;i}$.
Moreover, the eigenvalues corresponding to the singly-occupied sates ($|$\up$\rangle$ and $|$\down$\rangle$) gradually increase and exceed $\omega_{\textrm{\zero};i}$ and $\omega_{\textrm{\double};i}$.
In contrast to natural AP1roG orbitals, $\omega_{\textrm{\up};i}$ and $\omega_{\textrm{\down};i}$ have intermediate weights, which are similar to $\omega_{\textrm{\zero};i}$ and $\omega_{\textrm{\double};i}$.
Thus, optimization of the orbital basis within the AP1roG model reduces the contributions of $\omega_{\textrm{\up};i}$ and $\omega_{\textrm{\down};i}$ to the spectrum of $\rho_{i}$.
The corresponding eigenvalues of $\rho_{i,j}$ are shown in Figure~\ref{fig:rho2-ring}(a-ii).
AP1roG can accurately reproduce the dominant part of the spectrum of $\rho_{i,j}$ for squeezed hydrogen rings and around the equilibrium geometry (for $\omega_{\alpha;i,j}>0.01$).
For stretched metallic hydrogen rings, however, AP1roG fails to reproduce all $\omega_{\alpha;i,j}$ of the DMRG reference calculation, which are substantially over- or underestimated.
As observed for the optimized AP1roG natural orbitals, the eigenvalues corresponding to the singly-occupied states ($|$\up,\down$\rangle$, $|$\down,\up$\rangle$ etc.) gradually increase in magnitude when the hydrogen atoms are pulled apart.
Restricting the wavefunction to electron-pair states does not allow us to describe the orbital-pair correlations in stretched metallic hydrogen rings correctly.
This problem can be, at least partially, reduced if the orbital basis (and thus also the reference determinant) is optimized (cf.~Figures~\ref{fig:rho2-ring}(a-i) and \ref{fig:rho2-ring}(a-ii))

Similar observations can be made for the molecular H$_{14}$ ring. Figure~\ref{fig:mol1}(ii) shows $I_{i|j}$ and $s_i$ for different inter-molecular distances along the dissociation pathway of molecular H$_{14}$.
For canonical Hartree--Fock orbitals, AP1roG captures a larger fraction of the weak orbital-pair correlations ($I_{i|j}<10^{-2}$), but simultaneously underestimates the strong orbital-pair correlations.
As observed in the metallic hydrogen ring, the decay of $I_{i|j}({\rm AP1roG})$ agrees qualitatively well with the DMRG reference distribution (see Figure~\ref{fig:i12-ring}(b-ii)). Differences between AP1roG and DMRG are quantitative and amount to approximately one order of magnitude.
This behavior can be understood by analyzing the eigenvalues of $\rho_{i}$ and $\rho_{i,j}$, respectively, shown in Figures~\ref{fig:rho1-ring}(b-ii) and \ref{fig:rho2-ring}(b-ii).
If the orbitals are not optimized, AP1roG slightly over- and underestimates $\omega_{\alpha;i}$.
Note that the differences in $\omega_{\alpha;i}$ between AP1roG and DMRG are smaller in molecular than in metallic H$_{14}$.
In contrast to $\rho_{i}$, the eigenvalue spectrum of $\rho_{i,j}$ behaves differently when the hydrogen ring is stretched.
For increasing inter-molecular distances, AP1roG gradually underestimates $\omega_{\alpha;i,j}$ compared to the DMRG reference values, with differences amounting to one order of magnitude in the dissociation limit (that is, separated hydrogen molecules).
Note, however, that for $r_{\rm H-H}\le1.80$ bohr, the eigenvalue spectrum of $\rho_{i,j}$ can be accurately described by the AP1roG model, while for stretched hydrogen rings, the orbital basis needs to be optimized in order to reduce $\omega_{\alpha;i,j}$ attributed to the singly-occupied states.

Finally, we would like to comment on the influence of the size of the atomic orbital basis on orbital-pair correlations and orbital-entanglement.
We have performed additional calculations for metallic and molecular H$_{14}$ rings using the a correlation-consistent basis set of double-zeta quality (cc-pVDZ~\cite{dunning_b}) and stretched molecular geometries.
Most importantly, the overestimation of strong and intermediate orbital-pair correlations is not caused by the small basis set size used in our calculations.
Although increasing the atomic basis set to cc-pVDZ quality reduces the extent of overcorrelation, it does not completely eliminate the failures of AP1roG to provide reliable spectra of $\rho_{i}$ and $\rho_{i,j}$.
To remedy this problem, open-shell configurations have to be included in the wavefunction model.
The corresponding correlation diagrams are summarized in the Supporting Information.

\section{Conclusions}\label{sec:conclusions}
Wavefunctions constructed from electron-pair states, that is, so-called seniority-zero wavefunctions, are considered good models to describe strongly-correlated systems in condensed-matter physics and quantum chemistry.
However, most of the analysis presented so far was mainly based on energetic arguments or on evaluation of one-body correlation functions like occupation numbers.
In this work, we have presented an in-depth analysis of the correlations between the one-particle functions that are used to construct the geminals.
Specifically, we have scrutinized how accurately the AP1roG model can reproduce orbital-pair correlations and orbital-entanglement in the one-dimensional Hubbard model with periodic boundary conditions as well as in metallic and molecular hydrogen rings.

If the orbitals, and thus the reference determinant, are optimized, AP1roG can accurately describe the largest orbital-pair correlations in all investigated systems, misses, however, a large fraction of the weaker orbital-pair correlations.
In the strong correlation limit (large $U/t$ or in the vicinity of dissociation for metallic hydrogen rings), AP1roG considerably overestimates intermediate and weaker orbital-pair correlations ($I_{i|j}\le10^{-2}$) and results in a prolonged plateau of $I_{i|j}$.
This overcorrelation can be explained by the eigenvalue spectra of $\rho_{i}$ and $\rho_{i,j}$, which are used to determine the orbital-based correlation functions.
While in the weak correlation limit (small $U/t$ and metallic hydrogen around the equilibrium geometry), the eigenvalues corresponding to singly-occupied states $\omega_{\textrm{\up,\down};i,j}$, $\omega_{\textrm{\down,\up};i,j}$, etc.~are (orders of magnitudes) smaller than those corresponding to doubly-occupied or empty stats $\omega_{\textrm{\double,\double};i,j}$, $\omega_{\textrm{\zero,\zero};i,j}$, etc., their weights gradually increase when we approach the strong correlation regime.
Specifically, in the strong correlation limit, singly-occupied states become important and need to be included in the wavefunction model to accurately describe the spectrum of $\rho_{i}$ and $\rho_{i,j}$.
Specifically, states with unpaired electrons ($|$\down,\up$\rangle$, $|$\double,\up$\rangle$, $|$\double,\down$\rangle$, etc.), that is, $(n,s_z)=(2,0),(3,+\frac{1}{2}),(3,-\frac{1}{2}),\ldots$, have to be included into the wavefunction ansatz to properly describe orbital-pair correlations of order $10^{-2}$ or smaller.
It remains, however, ambiguous if the AP1roG model provides an accurate zero-order wavefunction (in the strong correlation limit) and if \textit{a posteriori} models, like perturbation theory or coupled-cluster-type corrections, provide enough flexibility to correct the (zero-order) orbital-pair correlations.
This is currenlty under investigation in our laboratory.

In the case of molecular hydrogen rings, AP1roG can accurately describe orbital-pair correlations along the dissociation pathway.
In contrast to metallic H$_{14}$, overestimation of orbital-pair correlations is negligible and only observable in the dissociation limit.
This overcorrelation can be attributed to small eigenvalues of $\rho_{i,j}$ corresponding to singly-occupied states, like $|$\down,\up$\rangle$, $|$\up,\down$\rangle$, etc.

If the one-particle functions are not optimized and the Hartree--Fock determinant is taken as reference determinant in the AP1roG ansatz, all orbital-pair correlations are smaller than the DMRG reference values.
Furthermore, AP1roG accurately reproduces the eigenvalue spectra of $\rho_{i}$ and $\rho_{i,j}$ ($\omega_{\alpha;i,j}\ge10^{-2}$) for small H--H distances, while it fails to reliably predict all eigenvalues $\omega_{\alpha;i,j}$ for stretched hydrogen rings.
Finally, we should note that for molecular geometries around the equilibrium structure ($r\approx1.80$ bohr), both $I_{i|j}$ and the eigenvalue spectra of $\rho_{i}$ and $\rho_{i,j}$ suggest that AP1roG provides accurate zero-order wavefunctions (with and without orbital optimization) where the missing orbital-pair correlations could be accurately modeled using \textit{a posteriori} approaches for weak electron correlation~\cite{Tamar-pCC,Boguslawski-lcc}.
A detailed analysis of orbital-pair correlations predicted by \textit{a posteriori} correlation models will be a subject of future publications.

\section{Acknowledgments}
K.B.~acknowledges financial support from a SONATA BIS grant of the National Science Centre, Poland (no.~2015/18/E/ST4/00584). 
P.T.~thanks the National Science Center Grant No.~DEC-2013/11/B/ST4/00771 and No.~DEC-2012/07/B/ST4/01347.
\"{O}.L. acknowledges financial support from the Hungarian Research Fund (OTKA K100908 and NN110360).

Calculations have been carried out using resources provided by Wroclaw Centre for Networking and Supercomputing (http://wcss.pl), grant No.~10105802.
\normalem
\bibliography{rsc} 
\end{document}